# Persistent Oscillations of X-ray Speckles: Pt (001) Step Flow

M. S. Pierce[1], D. C. Hennessy[1], K. C. Chang[1], V. Komanicky[1,2], J. Strzalka[3], A. Sandy[3], A. Barbour[1], and H. You[1]

[1] Materials Science Division, Argonne National Laboratory, Argonne Illinois 60439, USA

[2] Faculty of Sciences, Safarik University, Košice, 04001, Slovakia

[3] Advanced Photon Source, Argonne National Laboratory, Argonne Illinois 60439, USA

We observed well-defined oscillations of speckle intensities from Pt (001) surfaces at high temperatures, persisting for tens of minutes. We used a model of hex-reconstructed terraces to show that the coherent x-rays reflected from the terraces retain their phases relative to the illumination boundary and the observed oscillations come from surface dynamics due to "step-flow" motion. Our results demonstrate a possibility that x-ray speckles can be applied to monitor the real-time evolution of surfaces.

Sublimation of atoms from surfaces, essentially crystal growth in reverse, is a way to gain insight into the dynamic characteristics of surfaces. Various surfaces of Si[1,2,3] are the most extensively studied, and low energy electron microscopy (LEEM)[4,5] has been used to explore surface dynamics of Pt surfaces. Surface processes,[6,7] images,[8] and dynamics[9] have been studied by x-ray techniques. To date, however, the direct observation of step meandering or flow with x-rays has not been demonstrated. In this letter, we show that coherent surface x-ray scattering (CSXS) is directly applicable to measure such step motion and we apply it to study Pt (001) surface dynamics.

In reflection x-ray photon correlation spectroscopy, the exponentially-decaying or under-damping autocorrelations of speckle patterns represent the time evolution of the surface reconfiguration.[10,11,12] We find that the autocorrelations from the Pt (001) surface, however, can show persistent oscillations lasting many tens of cycles without explicit heterodyning. The observed oscillations in the autocorrelations are a unique feature of CSXS, qualitatively different from the oscillations of the incoherent surface scattering intensity seen in island growth.[13] While applications of CSXS to observe surface dynamics such as a step flow motion have long been understood to be possible,[13] our work represents an experimental realization of such idea.

We performed our experiments at 8ID of the Advanced Photon Source. A single bounce Si (111) monochromator was used to select ~$10^9$ incident photons with energy of 7.36 keV through a pair of 6×6 $\mu m^2$ slits. The base pressure inside the sample chamber was maintained at P ~ $10^{-6}$ Torr. An RF induction heater was used to control the sample temperature and the lattice constant was measured to determine the sample temperature.[14] Samples were polished to a nominally zero miscut, preannealed to bulk and surface mosaic widths of <0.1°, and cleaned in vacuum at ~ 1200 K for several hours immediately prior to data collection. A charge coupled device (CCD) with 20×20 μm$^2$ pixels was used for data collection, positioned 2.1 m from the sample. Speckle patterns were collected at the (001) anti-Bragg condition to maximize the sensitivity to the sample surface. Typically a data set was collected for 1-3 hours and the measured temperature drifted ≤1% during this period.

At high temperatures ≥ 1650 K, we observed that the hex domains of Pt (001) become highly-ordered[15] permitting us to access step-free regions within our illuminated area (6×30 μm$^2$). The step-free region produces a sharp single peak narrower than those at lower temperatures. Fig. 1 (a) shows the typical sharp peak. Remarkably, at temperatures close to the roughening transition ($T_R$=1830K),[15] we observed the sharp peak frequently splits to two peaks as shown in (b). The speckle pattern can then be observed to oscillate slowly between these two states over a long period of time. The time evolution of the oscillation (the whitened pixels) is plotted in (c). While the integrated intensity (not shown) is constant and exhibits no time-domain structure, the peak intensity at Y~0 shows clearly the oscillatory behavior with a ~40 sec period. Also, note that the side intensities (Y~±3) oscillate out of phase from the peak intensity. The images shown in Fig. 1 (a) and (b) were measured at 1808 K and consist of ~800 ph per 2 sec.

The oscillations are not expected in incoherent scattering; the scattering intensity is invariant on translation and remains constant even if the steps flow.[13] In coherent scattering, however, the invariance is broken by the step edge position relative to the boundary of the well-defined illuminated area. We considered a single-step flow model in which one step, at most, is present in the illuminated area. This will be the case if the step-step distance is the same as or larger than the width of the illuminated area as is our case. For example, the intensity at the center of the image (Y~0 or $\Delta\mathbf{q}$~0) at the (001) anti-Bragg condition will be the maximum value when the step is at either edge or outside the illumination area, and zero when the step is at the center. As the steps move, one at a time across the illuminated area, the intensity will oscillate as shown in Fig. 1 (d). Likewise, the intensity at Y~3 or $\Delta\mathbf{q}$~$\pi$/w (2w = illumination width) oscillates out of phase with respect to the intensity oscillation at $\Delta\mathbf{q}$~0.

While the ideal single-step model can produce the intensity oscillations, it can be modified to match the observed experimental data better. While the elongated the Y~0 peak can effectively simulated by the duration of the step-absence, the elongation of the side intensity can only be explained by the steps moving along the x-ray beam with an angle. The calculated intensity with an angle of ~7° is shown in (d) where the intensity oscillation is reasonably well reproduced. However, a close examination of the measured intensity (c) shows that i) the intensity of the upper side pixels is systematically stronger than that of the lower ones and ii) the peak positions of the lower pixels are ahead of the upper pixels. Both asymmetries in i) intensity and ii) time can be explained by a more realistic model that includes the hexagonally reconstructed terraces with the 19% relaxation.[16] We considered a model where the hex-reconstructed terrace must make a transition to the cubic (1×1) structure at the leading edge of its immediate upper terrace. The model is schematically shown in (f). Here, we allowed both the height relaxation and the density increase to continuously vary from the values for the hexagonal layer on a terrace to those for the first bulk layer. The simulation based on this model is shown in (e) where the transition region was set to 10% of a terrace width. We can surmise that the oscillation is from the step crossing while the systematic asymmetries result from the increased spacing and density of terraces.

To obtain quantitative dynamic information, we calculated the normalized autocorrelation,[7] $g_2(\Delta t)$ on a pixel by pixel basis, using only pixels of adequate intensity (≥3 ph/image). An example of the autocorrelations is given in Fig. 2(a). This example (and others not shown here) clearly exhibits persistent oscillations. The slow damping is a measure of how long on average it takes for the system to reconfigure due to thermal activities. Step meandering and hex domain fluctuation both contribute. The temperature dependence of the decorrelation rate is in itself interesting,[9] but here we focus on the oscillatory components in the autocorrelations. To isolate the oscillation in autocorrelation signal, Lorentzian fits to Fourier transformations of the autocorrelations for three different temperatures are shown in Fig. 2(b).

Below $T_R$=1830 K, the oscillation frequency becomes longer as the temperature is decreased as shown in Fig. 3. For 1650 < T < 1810 K, the frequencies are well described by Arrhenius function $f \propto e^{-E_A/k_BT}$ with an activation energy $E_A$ = 5.4(9) eV (black solid line). This value is similar to the heat of sublimation of Pt, 5.9 eV obtained from the known vapor pressure,[17] and thus a rather simple explanation of "step flow" can be given to the data. We expect atoms to readily leave the step edges, diffuse about the terraces for a time and come back to the step edges[18]. However, some of them, increasingly more as the temperature increases, leave the surface from the terraces and do not come back to the step edges. This net loss of atoms results in an overall, uniform step retraction. We believe that, in this temperature range, the number of steps does not change. Rather, the steps only retract faster as the temperature increases. We compared the frequency with the vapor-pressure data (open circles and dashed line)[17] assuming the unity sticking coefficient in Fig. 3. The slope of the black solid line is similar to that of dashed line indicating that the activation of the step edge retraction is ultimately governed by the sublimation energy. However, it is not surprising that they do not overlap. The sticking coefficient is likely to be significantly less than 1. Additionally, a single low index facet with large terraces is far from the statistical average surface where many other sublimation mechanisms are in operation.

The data points above 1810 K deviate to higher frequencies than the solid line. Since the energy for creating a step decreases to zero as T approaches $T_R$, the number of steps is expected to increase. In fact, it appears that the frequency diverges as T approaches $T_R$. Therefore, we assume that the step-density increase obeys a simple power law. Then, a modified version of Arrhenius equation, shown below, represents a heuristic attempt at describing the available data.

$$f \propto \left[1 + a_0(T_R - T)^{-\alpha}\right] e^{-E_A/k_B T} \quad (1)$$

This equation is plotted as the red line in Fig. 3 with an exponent value of α≈1.4(3) and a proportionality constant $a_0$. Although the divergence of the step creation towards the $T_R$ is plausible, no theoretical justification for the exponent is given and the value remains qualitative.

In summary, we have observed persistent oscillations in the speckle intensities and the corresponding autocorrelations. The oscillations are found to be consistent with the step-flow motion of the Pt (001) surface at high temperature due to sublimation. Our results demonstrate that CSXS can yield information about uniform motions of surface features in addition to the decorrelation dynamics. We believe that CSXS techniques such as this can find many useful *in* *situ* applications for systems buried under electrolytes, high-pressure gas, or other extreme environments, particularly as modern x-ray sources and timing techniques continue to advance.

This work at ANL and use of the Advanced Photon Source were supported by the U.S. DOE, Office of Basic Energy Sciences, under Contract No. DE-AC02-06CH11357. The work at SU was supported by Slovak grant VEGA 1/0138/10 and VVCE-0058-007.

**Figures:**

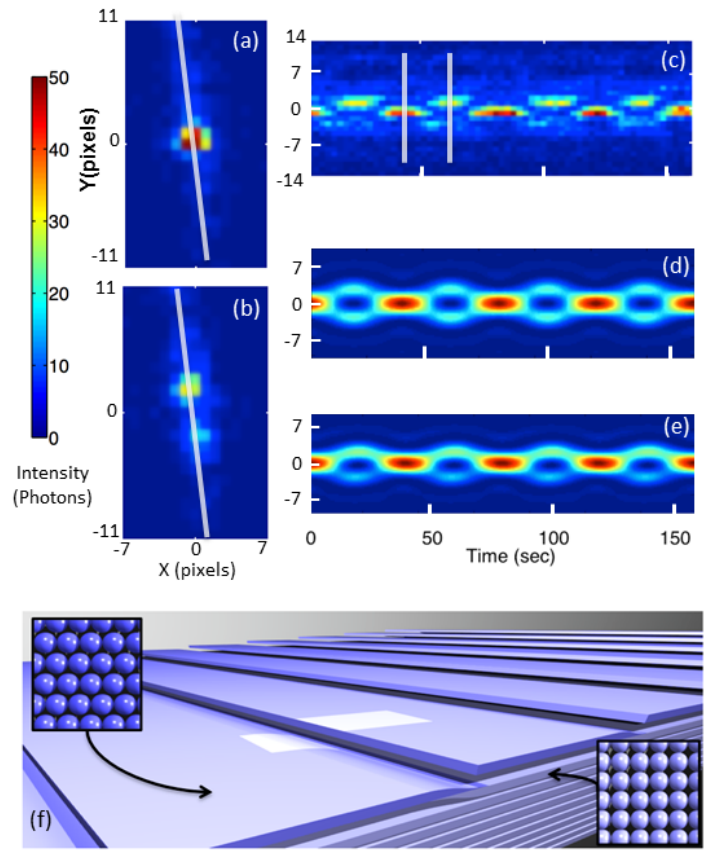

Figure 1. (a) A CCD image where *x* is the direction across the beam and *y* is in the 2θ (≈25º) direction. (b) The same region 20 sec later. (c) The highlighted intensity from panels (a) and (b) vs. time where the vertical lines indicate the time at which (a) and (b) were recorded. Calculated intensity vs. time for the single step model (d) without and (e) with the reconstructed terraces. (f) A model of reconstructed terraces. The rectangular illumination area is shown in white at the center and ball models of the hexagonal and cubic lattices are shown as insets.

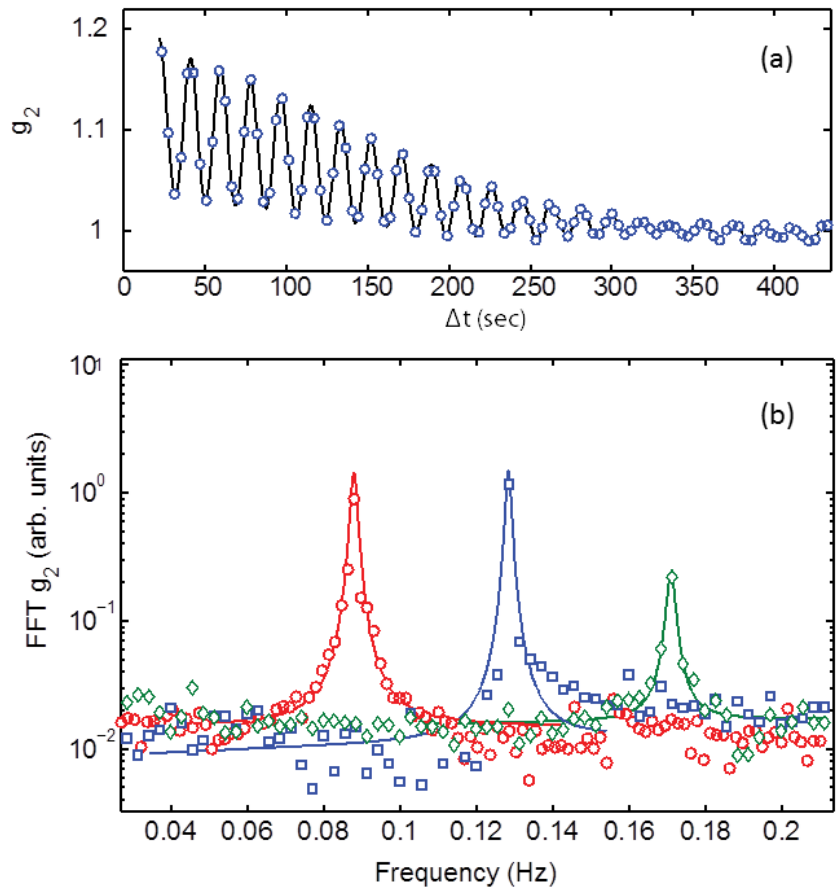

Figure 2. (a) Autocorrelation $g_2$ measured at 1822 K. (b) Fourier transformations of $g_2$ measured at 1822, 1825, and 1827 K.

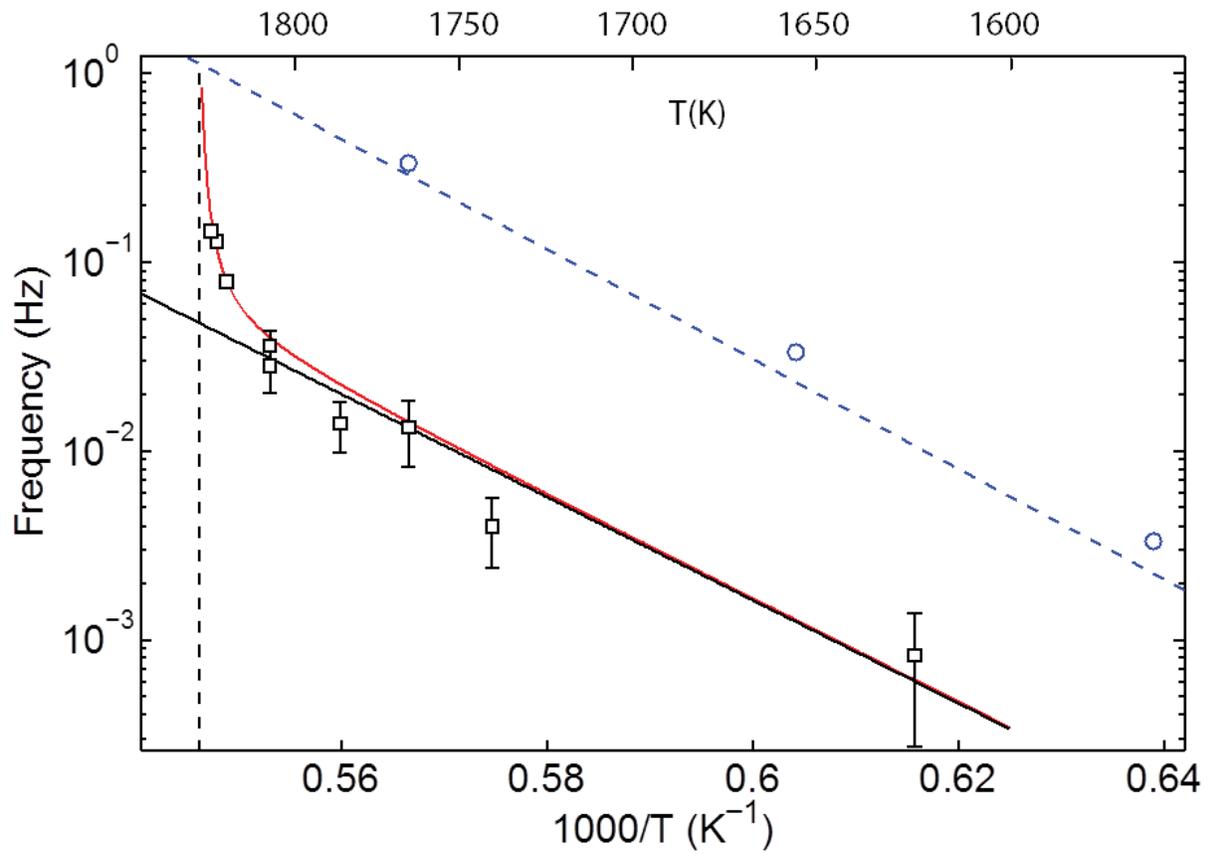

Figure 3. Frequency vs. inv. temperature. The black and red solid lines are fits to the data, and the blue circles and dashed line are from the Pt vapor pressure.